\documentclass[times]{nmeauth}

\usepackage{moreverb}
\usepackage[section]{placeins}
\usepackage{float}

\usepackage[dvips,colorlinks,bookmarksopen,bookmarksnumbered,citecolor=red,urlcolor=red]{hyperref}

\newcommand\BibTeX{{\rmfamily B\kern-.05em \textsc{i\kern-.025em b}\kern-.08em
T\kern-.1667em\lower.7ex\hbox{E}\kern-.125emX}}

\def\volumeyear{2015}

\usepackage{kurz}
\usepackage{amssymb}
\usepackage{mathrsfs}
\usepackage{multirow}    
\usepackage{amsmath}
\usepackage{caption}  
\usepackage{stmaryrd}
\usepackage{rotating}
\usepackage{setspace}

\usepackage[numbers,sort&compress]{natbib}

\usepackage{alphalph}
\makeatletter
\newalphalph{\fnsymbolwrap}[wrap]{\@fnsymbol}{}
\makeatother

\begin{document}



\title{Global cracking elements: a novel tool for Galerkin-based approaches simulating quasi-brittle fracture}

\author{Yiming Zhang\affil{a}\corrauth, Herbert A. Mang\affil{b,c,d}\corrauth}


\address{\affilnum{a}School of Civil and Transportation Engineering, Hebei University of Technology, Xiping Road 5340, 300401~Tianjin,~P.R.China \break 
	\affilnum{b}Department of Geotechnical Engineering, Tongji University, Siping Road 1239, 200092~Shanghai,~P.R.China \break 
	\affilnum{c}State Key Laboratory for Disaster Reduction in Civil Engineering, Tongji University, Siping Road 1239, 200092~Shanghai,~P.R.China \break 
	\affilnum{d}Institute for Mechanics of Materials and Structures (IMWS), Vienna University of Technology,Karlsplatz 13/202, 1040 Vienna, Austria}

\corraddr{Herbert A. Mang and Yiming Zhang\\
	 \mbox{~~~~~~~~~~~~~~~~~~~~~~~Email:} Herbert.Mang@tuwien.ac.at, Yiming.Zhang@hebut.edu.cn}

\doublespacing

\large

\begin{abstract}
Following the so-called Cracking Elements Method (CEM), recently presented in \cite{Yiming:14,Yiming:16}, we propose a novel Galerkin-based numerical approach for simulating quasi-brittle fracture, named Global Cracking Elements Method (GCEM).  For this purpose the formulation of the original CEM is reorganized.  The new approach is embedded in the standard framework of the Galerkin-based Finite Element Method (FEM), which uses disconnected element-wise crack openings for capturing crack initiation and propagation.  The similarity between the proposed Global Cracking Elements (GCE) and the standard 9-node quadrilateral element (Q9) suggests a special procedure: the degrees of freedom of the center node of the Q9, originally defining the displacements, are ``borrowed" to describe the crack openings of the GCE.  The proposed approach does not need remeshing, enrichment, or a crack-tracking strategy, and it avoids a precise description of the crack tip.  Several benchmark tests provide evidence that the new approach inherits from the CEM most of the advantages.  The numerical stability and robustness of the GCEM are better than the ones of the CEM.  However, presently only quadrilateral elements with nonlinear interpolations of the displacement field can be used.

\end{abstract}


\maketitle

\vspace{-6pt}

\section{Introduction}
Fracture of quasi-brittle materials is accompanied by a fast strain softening process, resulting in a localized failure zone of very small size \cite{deBorst:07,Wujianying:02}.  In the last decades, different approaches were presented.  They are made of a continuum \cite{Wujianying:03,Wujianying:04}, discrete \cite{Areias:02,Areias:06} or particle/meshless \cite{Zhuang:01,Zhuang:03,Rabczuk2007743} based framework, embedded either in classical fracture mechanics, or in continuum damage mechanics, or in an equivalent-type theory such as peridynamics and lattice models \cite{Silling:01,HAN2016453,Zhaogaofeng2017}.

For evaluation of the ease of implementation, the reliability, robustness, and the efficiency of numerical methods for simulation of quasi-brittle fracture, the following criteria are listed in form of questions with ``yes" or ``no" answers: 
\begin{enumerate}
	\item
	Does the new method stay far outside of a conventional continuum-based framework?
	\item
	Does it provide results that strongly depend on the discretization?
	\item
	Does it require reprocessing such as remeshing?
	\item
	Does it introduce extra degrees of freedom other than displacements?
	\item
	Does it need to predict the crack path (position as well as orientation)?
	\item
	Does it require more computing effort than a traditional FEM-based method such as the smeared-crack apprach?
\end{enumerate}
A good method is one with predominantly negative answers to these questions.  In this paper, a novel numerical approach is presented, with ``No" as the answer to all of the above questions.  It is named the Global Cracking Elements Method (GCEM).  Its characteristics are as follows:
\begin{enumerate}
   \item
   The GCEM is developed on the basis of the original cracking elements method (CEM), presented in \cite{Yiming:14,Yiming:16,Yiming:11}.  It is embedded in the conventional continuum-based framework.  It may be considered as a conventional FEM approach with a special type of elements.
   \item
   Unlike the smeared-crack approach \cite{Cervera:10}, the discretization dependency of the GCEM is negligible.  To demonstrate this advantage, all of the numerical examples in this work are characterized by irregular meshes.  Besides, the mesh used in the GCEM can be relatively coarse. 
   \item
   For putting it more precisely, the GCEM is a type of Strong-Discontinuity-embedded Approach (SDA) \cite{Oliver:11,Saloustros:01,Dias-da-Costa:02}, which obviously does not need remeshing.  
   \item
   Unlike phase field methods \cite{Miehe:01,deBorst:04}, extended finite element methods (XFEM) \cite{SongandBelytschko,Chau-Dinh2012242}, or numerical manifold methods (NMM) \cite{Wuzhijun2012,ZhengHong:04,Wuzhijun2017}, the GCEM does not require a precise description of the stress state at crack tips and does not need extra degrees of freedom\footnote{Nonetheless, a small trick is used insofar as the degrees of freedom of the center node of the Q9 are ``borrowed" for indicating crack openings, see the following sections for details.}.  
   \item
   Unlike the traditional SDA and the XFEM, the GCEM does not need a crack-tracking strategy \cite{Dumstorff:01,Saloustros2018}.  Disconnected element-wise cracking segments, passing through the elements, are used for representing crack paths.  The orientation of the crack is determined locally, providing \emph{self-propagating} cracks \cite{Yiming:14}.  
   \item
   The global cracking elements used in the framework of the GCEM, are formally similar to the 9-node quadrilateral elements.  They provide a symmetric and sparse stiffness matrix \cite{Belytschko:01,Yiming:11}.  Thus the numerical efficiency is reasonably high.
\end{enumerate}

The main features of the GCEM are as follows:
\begin{itemize} 
	\item
	The designed special type of global cracking elements is formally similar to the Q9 element.  The shape functions, the local stiffness matrix, and the residuals will be provided in this work.
	\item
	The domain is initially discretized by Q9 elements.  Once a crack appears, the standard Q9 element will become a global cracking element by ``borrowing" the displacement degrees of freedom to indicate local crack openings.
	\item
	Since the crack openings become global unknowns, the GCEM is numerically more stable and more efficient than the CEM.  Fewer iteration steps than in the CEM are needed.  At the same time, the self-propagating property of cracking elements is maintained. 
\end{itemize}

The paper is organized as follows: In Section~\ref{sec:form}, the adopted mixed-mode softening law, the kinematics, and the designed cracking element are described.  Numerical examples are presented in Section~\ref{sec:nurm}, including an L-shaped panel test and disk tests with initial slots. In all examples, irregular meshes, with different sizes of the elements, are used.  Section~\ref{sec:conc} contains concluding remarks.

\section{Formulation}
\label{sec:form}

\subsection{Traction-separation law}
\begin{figure}[htbp]
	\centering
	\includegraphics[width=0.8\textwidth]{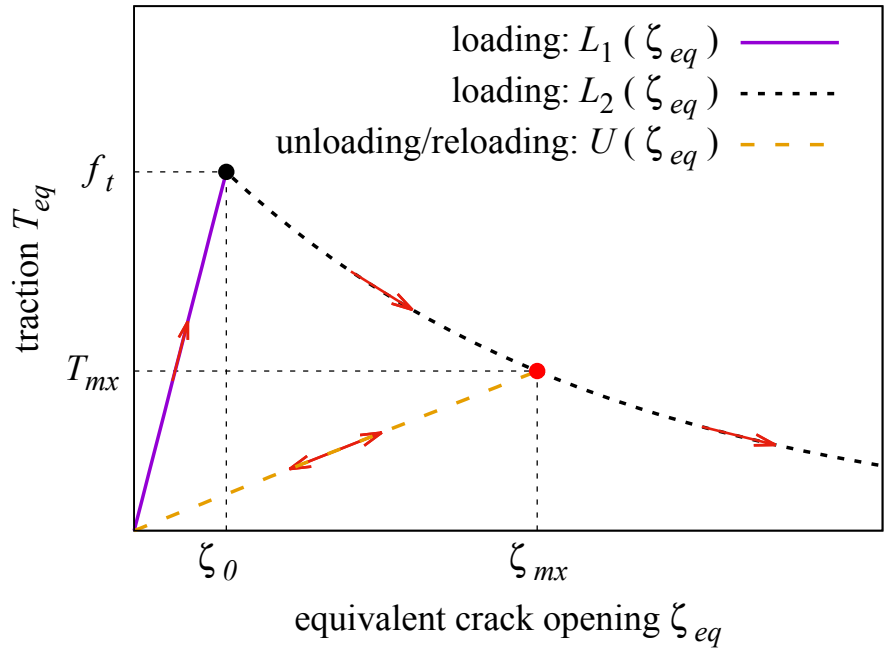}
	\caption{Traction-separation curve $T_{eq}\left(\zeta_{eq} \right)$}
	\label{fig:traction}
\end{figure}

A mixed mode traction-separation law \cite{Meschke:01,Yiming:15} is used in this work.  With $\zeta_n$ and $\zeta_t$ denoting the crack openings in the normal and the parallel direction, respectively, the equivalent crack opening is defined as
\begin{equation}
\zeta_{eq}=\sqrt{\zeta_n^2+\zeta_t^2}.
\label{eq:zetaeq}
\end{equation}
The traction components in the normal and the parallel direction, i.e. $T_n$ and $T_t$, are obtained as
\begin{equation}
\begin{aligned}
&T_n=T_{eq}\frac{\zeta_n}{\zeta_{eq}},~ T_t=T_{eq}\frac{\zeta_t}{\zeta_{eq}}\\
&\mbox{with }\\
&T_{eq}\left(\zeta_{eq} \right)=\left\{\begin{array}{ll}
L_1\left(\zeta_{eq} \right)=\cfrac{f_t}{\zeta_{0}}\ \zeta_{eq},& \mbox{loading}, \zeta_{eq}\leq \zeta_0\\
L_2\left(\zeta_{eq} \right)=f_t \ \mbox{exp}\left[-\cfrac{f_t\left(\zeta_{eq}-\zeta_0\right)}{G_f-G_{f,0}}\right],& \mbox{loading}, \zeta_{eq}>\zeta_0\\
U\left(\zeta_{eq} \right)=\cfrac{T_{mx}}{\zeta_{mx}}\ \zeta_{eq},& \mbox{unloading/reloading},
\end{array}\right.
\end{aligned}
\label{eq:Traction}
\end{equation}
where $f_t$ is the uniaxial tensile strength, $G_f$ is the fracture energy, $G_{f,0}$ is the threshold value of $G_f$, with $G_{f,0}=0.001\ G_f$ assumed herein, $\zeta_0$ is the threshold opening, given as $\zeta_0=2\  G_{f,0}/f_t$, and $\zeta_{mx}$ denotes the maximum opening, which the crack has ever experienced.  Its value is updated at the end of every loading step if $\zeta_{mx}>\zeta_0$.  $T_{mx}=L_2\left(\zeta_{mx}\right) $ is the corresponding traction.  Figure~\ref{fig:traction} shows $T_{eq}\left(\zeta_{eq} \right)$.

From equations~\ref{eq:zetaeq} and~\ref{eq:Traction}, the following relations can be obtained
\begin{equation}
\begin{aligned}
&\mathbf{D}=\left[\begin{array}{cc}
{\partial T_n}/{\partial \zeta_n}&{\partial T_n}/{\partial \zeta_t}\\
{\partial T_t}/{\partial \zeta_n}&{\partial T_t}/{\partial \zeta_t}\\
\end{array}\right]=\\
&\left\{\begin{array}{ll}
\cfrac{f_t}{\zeta_0}\left[\begin{array}{cc}
1&0\\
0&1\\
\end{array}\right],& \mbox{loading}, \zeta_{eq}\leq \zeta_0\\
\\
-\cfrac{T_{eq}}{\zeta_{eq}^2}\left[\begin{array}{cc}
\cfrac{\zeta_n^2}{\zeta_{eq}}+\cfrac{f_t\ \zeta_n^2}{{G_f-G_{f,0}}}-\zeta_{eq}&
\cfrac{\zeta_n\ \zeta_t}{\zeta_{eq}}+\cfrac{f_t\ \zeta_n\ \zeta_t}{{G_f-G_{f,0}}}\\
\cfrac{\zeta_n\ \zeta_t}{\zeta_{eq}}+\cfrac{f_t\ \zeta_n\ \zeta_t}{{G_f-G_{f,0}}}&
\cfrac{\zeta_t^2}{\zeta_{eq}}+\cfrac{f_t\ \zeta_t^2}{{G_f-G_{f,0}}}-\zeta_{eq}\\
\end{array}\right],& \mbox{loading}, \zeta_{eq}>\zeta_0\\
\\
\cfrac{T_{mx}}{\zeta_{mx}}\left[\begin{array}{cc}
1&0\\
0&1\\
\end{array}\right],& \mbox{unloading/reloading}.
\end{array}\right.
\label{eq:dTraction}
\end{aligned}
\end{equation}
These relations will be used for computation of the stiffness matrix and the residual of the cracking element.

\subsection{Kinematics}
\begin{figure}[htbp]
	\centering
	\includegraphics[width=0.85\textwidth]{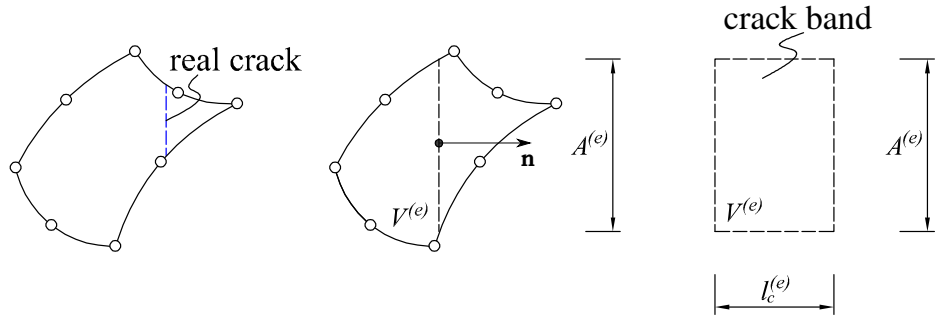}
	\caption{Determination of the value of $l_c$ of element $e$, with the volume $V^{(e)}$ and the effective cracking area $A^{(e)}$ (based on a parallel crack passing through the center point)}
	\label{fig:lc}
\end{figure}

The deduction of the displacement field $\mathbf{u}(\mathbf{x})$ and of the corresponding strain field $\boldsymbol{\varepsilon}(\mathbf{x})$ of the cracking element can be found in \cite{Yiming:11,Yiming:14}.  The result for $\mathbf{u}(\mathbf{x})$ is given as
\begin{equation}
\mathbf{u}(\mathbf{x}) = \bar{\mathbf{u}}(\mathbf{x}) + [H_s(\mathbf{x})-\varphi(\mathbf{x})] \llbracket u \rrbracket,
\label{eq:dispu}
\end{equation}
where $\bar{\mathbf{u}}(\mathbf{x})$ is the regular part, $H_s(\mathbf{x})$ is the Heaviside function across the discontinuity surface, with $H_s(\mathbf{x})=1$ on one side of the subdomain and $H_s(\mathbf{x})=0$ on the other side, and $\varphi(\mathbf{x})$ is a differentiable function with $\varphi(\mathbf{x})\in \left[0,1\right]$.  The result for $\bar{\boldsymbol{\varepsilon}}(\mathbf{x})$ is obtained as
\begin{equation}
\begin{array}{cccc}
\bar{\boldsymbol{\varepsilon}}(\mathbf{x})=&\underbrace{\nabla^S \bar{\mathbf{u}}(\mathbf{x})}&-&\underbrace{\left[(\mathbf{n} \otimes \nabla \varphi)^S \zeta_n(\mathbf{x})+(\mathbf{t} \otimes \nabla \varphi)^S \zeta_t(\mathbf{x})\right]},\\
&\mbox{total strain } \widehat{\boldsymbol{\varepsilon}}&&\mbox{enhanced strain } \widetilde{\boldsymbol{\varepsilon}}
\end{array}
\label{eq:EAS}
\end{equation}
where, unit vectors corresponding to $\zeta_n$ and $\zeta_t$ are denoted as $\mathbf{n}$ and $\mathbf{t}$, respectively.  
In the presented symmetric formulation\footnote{In this approach $\varphi\left(\mathbf{x}\right)$ does not have to be known, while in the XFEM and in some other types of SDA $\varphi\left(\mathbf{x}\right)=\sum N_i \varphi_i$ is assumed.  Therefore, the proposed approach is easier to implement, and it has fewer assumptions.}, $\nabla \varphi \parallel \mathbf{n}$.  Then, based on energy conservation \cite{Yiming:11}, $\nabla \varphi={\mathbf{n}}\ / \ {l_c}$ is obtained, with $l_c=V/A$.  In this formula, $V$ is the volume of the cracking element and $A$ is the area of an equivalent crack surface, i.e. the surface of a parallel crack, passing through the center point of the element, as shown in Figure~\ref{fig:lc}.  Hereby, $l_c$ corresponds to the classical characteristic length.  It only depends on the shape of the element and on the orientation of the crack, but not on its position.  

For the FEM approximation of the 8-node quadrilateral element (Q8), Eq.~\ref{eq:EAS} is obtained as
\begin{equation}
\begin{array}{cccc}
\bar{\boldsymbol{\varepsilon}}^{(e)}\approx&\underbrace{\sum^{8}_{i=1}\left(\nabla N^{(e)}_i \otimes \mathbf{u}_i\right)^S}&-&\underbrace{\frac{1}{\ l_c^{(e)} \ }\left[(\mathbf{n}^{(e)} \otimes  {\mathbf{n}}^{(e)}) \zeta_n^{(e)}+(\mathbf{n}^{(e)} \otimes  \mathbf{t}^{(e)})^S \zeta^{(e)}_t\right]},\\
&\widehat{\boldsymbol{\varepsilon}}^{(e)}&&\widetilde{\boldsymbol{\varepsilon}}^{(e)}
\end{array}
\label{eq:EAS_e}
\end{equation}
where $\left( \cdot \right)^S$ denotes the symmetric part of the tensor \cite{Mosler:01} and $\left( \cdot \right)^{(e)}$ is the value of the respective quantity of the Q8 element $e$.  E.g., $N_i^{(e)}$ is a shape function.

Furthermore, in \cite{Yiming:11,Yiming:14,Yiming:16} an idea referred to as ``\emph{center representation}", is presented.  Once a crack appears in an element, the stress/strain state at the center Gauss point is used for representation of the stress/strain state of the whole element.  Considering the equivalence of forces of discrete and embedded models \cite{Yiming:11}, the following relations exist not just at the center point, but in the whole cracking element:
\begin{equation}
\begin{aligned}
&\boldsymbol{\sigma}^{(e)}=\mathbb{C}^{(e)}:\bar{\boldsymbol{\varepsilon}}^{(e)}\ \ \ \mbox{and}\\
&\left\{\begin{array}{l}
\left(\mathbf{n}^{(e)}\otimes\mathbf{n}^{(e)}\right):\boldsymbol{\sigma}^{(e)}-T_n^{(e)}=0\\
\left(\mathbf{n}^{(e)}\otimes\mathbf{t}^{(e)}\right)^S:\boldsymbol{\sigma}^{(e)}-T_t^{(e)}=0
\end{array}\right. ,
\label{eq:equcenter}
\end{aligned}
\end{equation}
where $\mathbb{C}^{(e)}$ is the elasticity tensor.

\subsection{Global Cracking Element}
In this subsection, the Voigt notation is used for representation of second- and fourth-order tensors with corresponding vector and matrix forms \cite{Heinwein:01}. For the Q8 element, the displacement vector is given as $\mathbf{U}^{(e)}=\left[\mathbf{u}^{(e)}_1 \cdots \mathbf{u}^{(e)}_8\right]^T$.  The $\mathbf{B}$ matrix is given as
\begin{equation}
\begin{aligned}
&\mathbf{B}^{(e)}=
\left[ {\begin{array}{ccc}
	\mathbf{B}^{(e)}_1&\cdots&\mathbf{B}^{(e)}_8\\
	\end{array} } \right],\\
&\mbox{where}\\
&\mathbf{B}^{(e)}_i=\left[\begin{array}{ccc}
\cfrac{\partial N_i^{(e)}}{\partial x} &0\\
0&\cfrac{\partial N_i^{(e)}}{\partial y} \\
\cfrac{\partial N_i^{(e)}}{\partial y}&\cfrac{\partial N_i^{(e)}}{\partial x}
\end{array}
\right],\ i=1 \cdots 8.\\
\label{eq:B}
\end{aligned}
\end{equation} 
The unit vectors $\mathbf{n}^{(e)}$ and $\mathbf{t}^{(e)}$ are given as $\mathbf{n}^{(e)}=\left[n^{(e)}_x, n^{(e)}_y\right]^T$ and $\mathbf{t}^{(e)}=\left[t^{(e)}_x, t^{(e)}_y\right]^T$.  Then, based on Eq.~\ref{eq:EAS_e}, a special matrix $\mathbf{B}^{(e)}_\zeta$ is introduced as follows:
\begin{equation}
\mathbf{B}^{(e)}_\zeta=\frac{-1}{\ l_c^{(e)} \ }
\left[ {\begin{array}{c}
	\mathbf{n}^{(e)}\otimes\mathbf{n}^{(e)}\\
	\left(\mathbf{n}^{(e)}\otimes\mathbf{t}^{(e)}\right)^S
	\end{array} } \right]^T
=\frac{-1}{\ l_c^{(e)} \ }
\left[ {\begin{array}{cc}
	n^{(e)}_x\cdot n^{(e)}_x&n^{(e)}_x\cdot t^{(e)}_x\\
	n^{(e)}_y\cdot n^{(e)}_y&n^{(e)}_y\cdot t^{(e)}_y\\
	2\ n^{(e)}_x\cdot n^{(e)}_y&n_x\cdot t^{(e)}_y+n^{(e)}_y\cdot t^{(e)}_x
	\end{array} } \right].
\label{eq:Bz}
\end{equation} 
Thus, $\widetilde{\boldsymbol{\varepsilon}}^{(e)}=-\mathbf{B}^{(e)}_\zeta \  \boldsymbol{\zeta}^{(e)}$, where $\boldsymbol{\zeta}^{(e)}=\left[\zeta_n^{(e)}\ \  \zeta_t^{(e)}\right]^T$.  With $\mathbf{B}^{(e)}_\zeta$, Eq.~\ref{eq:equcenter} can be rewritten as follows:

\begin{equation}
-l_c\left(\mathbf{B}^{(e)}_\zeta\right)^T\boldsymbol{\sigma}^{(e)}-\left[ {\begin{array}{c}
	T_n^{(e)}\\
	T_t^{(e)}
	\end{array} } \right]=\mathbf{0}.
\label{eq:equcentermatrix}
\end{equation}
Next, for convenience we assume that $\mathbf{B}^{(e),1}$ denotes $\mathbf{B}^{(e)}$ for the center Gauss point.  Thus, Eq.~\ref{eq:EAS_e} gives
\begin{equation}
\bar{\boldsymbol{\varepsilon}}^{(e)}=\bar{\boldsymbol{\varepsilon}}^{(e),1}=\left[ {\begin{array}{cc}
	\mathbf{B}^{(e),1}&	\mathbf{B}^{(e)}_\zeta
	\end{array} } \right] \left[ {\begin{array}{c}
	\mathbf{U}^{(e)}\\
	\boldsymbol{\zeta}^{(e)}
	\end{array} } \right].
\label{eq:EASQ8}
\end{equation}
The elastic energy $E^{(e)}$ is obtained as
\begin{equation}
E^{(e)}=\frac{1}{2}\int \boldsymbol{\sigma}^{e}\cdot\bar{\boldsymbol{\varepsilon}}^{(e)}d(e)=\frac{V^{(e)}}{2}\ \left(\bar{\boldsymbol{\varepsilon}}^{(e)}\right)^T\ \mathbf{C}^{(e)}\ \bar{\boldsymbol{\varepsilon}}^{(e)},
\label{eq:Eg}
\end{equation}
where $\int(\cdot)d(e)$ is the integral of $\boldsymbol{\sigma}^{e}\cdot\bar{\boldsymbol{\varepsilon}}^{(e)}$ in the element $e$ and $\mathbf{C}^e$ denotes the matrix form of $\mathbb{C}^{(e)}$.  The elastic energy is used for checking whether the equilibrium iteration by means of the Newton-Raphson (N-R) method converges, see the next section for the details.

For the iteration step $l$ within load step $i$ in the framework of the N-R iteration \cite{Yiming:14,Yiming:15}, the following relation is obtained:
\begin{equation}
\begin{array}{cccc}
\left[ {\begin{array}{c}
	\mathbf{U}_{i,l}^{(e)}\\
	\boldsymbol{\zeta}_{i,l}^{(e)}
	\end{array} } \right]=
&\underbrace{
	\left[ {\begin{array}{c}
		\mathbf{U}^{(e)}_{i-1}\\\boldsymbol{\zeta}^{(e)}_{i-1}
		\end{array} } \right]+
	\left[ {\begin{array}{c}
		\Delta\mathbf{U}^{(e)}_{l-1}\\\Delta\boldsymbol{\zeta}^{(e)}_{l-1}
		\end{array} } \right]}&+&\underbrace{
	\left[ {\begin{array}{c}
		\Delta\Delta\mathbf{U}^{(e)}\\\Delta\Delta\boldsymbol{\zeta}^{(e)}
		\end{array} } \right]},\\
&\mbox{known}&&\mbox{unknown}\\
\end{array}
\label{eq:UdU2}
\end{equation}
where $\Delta\left(\cdot\right)$ denotes an increment with respect to the corresponding value at the preceding load step, $i-1$, and  $\Delta\Delta\left(\cdot\right)$ stands for an increment with respect to the value at the last N-R iteration step, $l-1$.  The N-R iteration process was described in \cite{Yiming:16}.

The balance equation $\nabla \boldsymbol{\sigma}=\mathbf{F}$ for quasi-static loading, where $\mathbf{F}$ denotes the loading force, and making use of the equivalence relations according to Eq.~\ref{eq:equcentermatrix}, after linearization the global balance equation is obtained as follows:
\begin{equation}
\begin{aligned}
&\mathbf{K}_s^{(e)}
\left[ {\begin{array}{c}
	\Delta\Delta\mathbf{U}\\\Delta\Delta\boldsymbol{\zeta}
	\end{array} } \right]=
\left[ {\begin{array}{c}
	\mathbf{F}\\\mathbf{T}
	\end{array} } \right]-\mathbf{K}^{(e)}
\left[ {\begin{array}{c}
	\mathbf{U}^{(e)}_{i-1}+\Delta\mathbf{U}^{(e)}_{l-1}\\\boldsymbol{\zeta}^{(e)}_{i-1}+\Delta\boldsymbol{\zeta}^{(e)}_{l-1}
	\end{array} } \right],\\
&\mbox{with}\\
&\mathbf{K}^{(e)}=
\left[ {\begin{array}{cc}
	\int \left(\mathbf{B}^{(e)}\right)^T \mathbf{C}^e \ \left(\mathbf{B}^{(e),1}\right) d(e) &\int \left(\mathbf{B}^{(e)}\right)^T \mathbf{C}^e \ \mathbf{B}_\zeta^{(e)} d(e) \\
	-l_c\left(\mathbf{B}^{(e)}_\zeta\right)^T\mathbf{C}^e \ \left(\mathbf{B}^{(e),1}\right) &-l_c\left(\mathbf{B}^{(e)}_\zeta\right)^T\mathbf{C}^e \ \mathbf{B}^{(e)}_\zeta\\
	\end{array} } \right]\\
&\mbox{and}\\
&\mathbf{K}_s^{(e)}=\mathbf{K}^{(e)}-
\left[ {\begin{array}{cc}
	\mathbf{0}&\mathbf{0}\\
	\mathbf{0}&\mathbf{D}^{(e)}\\
	\end{array} } \right],
\label{eq:Ksub}
\end{aligned}
\end{equation} 
where, $\mathbf{T}=\mathbf{T}\left(\boldsymbol{\zeta}^{(e)}_{i-1}+\Delta\boldsymbol{\zeta}^{(e)}_{l-1}\right)=\left[T_n\ \  T_t\right]^T$.  Eq.~\ref{eq:Ksub} is not useful for numerical analysis, because neither $\mathbf{K}_s^{(e)}$ nor $\mathbf{K}^{(e)}$ is symmetric.  As a remedy of this problem, Eq.~\ref{eq:Ksub} is slightly changed to
\begin{equation}
\begin{aligned}
&\mathbf{K}_{s,new}^{(e)}
\left[ {\begin{array}{c}
	\Delta\Delta\mathbf{U}\\\Delta\Delta\boldsymbol{\zeta}
	\end{array} } \right]=
\left[ {\begin{array}{c}
	\mathbf{F}\\-{V^{(e)}} \ / \ {l_c^{(e)}}\ \mathbf{T}
	\end{array} } \right]-\mathbf{K}_{new}^{(e)}
\left[ {\begin{array}{c}
	\mathbf{U}^{(e)}_{i-1}+\Delta\mathbf{U}^{(e)}_{l-1}\\\boldsymbol{\zeta}^{(e)}_{i-1}+\Delta\boldsymbol{\zeta}^{(e)}_{l-1}
	\end{array} } \right],\\
&\mbox{with}\\
&\mathbf{K}_{new}^{(e)}=
\left[ {\begin{array}{cc}
	\int \left(\mathbf{B}^{(e)}\right)^T \mathbf{C}^e \ \left(\mathbf{B}^{(e),1}\right) d(e) &\int \left(\mathbf{B}^{(e)}\right)^T \mathbf{C}^e \ \mathbf{B}_\zeta^{(e)} d(e) \\
	V^{(e)} \left(\mathbf{B}^{(e)}_\zeta\right)^T\mathbf{C}^e \ \left(\mathbf{B}^{(e),1}\right) &V^{(e)} \left(\mathbf{B}^{(e)}_\zeta\right)^T\mathbf{C}^e \ \mathbf{B}^{(e)}_\zeta\\
	\end{array} } \right]\\
&\mbox{and}\\
&\mathbf{K}_{s,new}^{(e)}=\mathbf{K}_{new}^{(e)}+
\left[ {\begin{array}{cc}
	\mathbf{0}&\mathbf{0}\\
	\mathbf{0}&{V^{(e)}} \ / \ {l_c} \ \mathbf{D}^{(e)}\\
	\end{array} } \right].
\label{eq:Ksub2}
\end{aligned}
\end{equation} 
However, also $\mathbf{K}_{s,new}^{(e)}$ is not symmetric.  Moreover, this matrix has many zero elements, rendering the iteration numerically unstable.  Fortunately, $\mathbf{K}_{s,new}^{(e)}$ can be replaced by the following symmetric matrix:
\begin{equation}
\mathbf{K}_{sym}^{(e)}=\int \left[ {\begin{array}{cc}
	\mathbf{B}^{(e)}&	\mathbf{B}^{(e)}_\zeta
	\end{array} } \right]^T \mathbf{C}^e \left[ {\begin{array}{cc}
	\mathbf{B}^{(e)}&	\mathbf{B}^{(e)}_\zeta
	\end{array} } \right] d(e) +
\left[ {\begin{array}{cc}
	\mathbf{0}&\mathbf{0}\\
	\mathbf{0}&{V^{(e)}} \ / \ {l_c} \ \mathbf{D}^{(e)}\\
	\end{array} } \right].
\label{eq:Ksym}
\end{equation} 
It is worth mentioning that the matrix $\mathbf{K}_{new}^{(e)}$, see Eq.~\ref{eq:Ksub2}, appearing as the final term on the right-hand side of Eq.~\ref{eq:Ksub2} remains the same.  Otherwise serious stress locking would occur (see Appendix A of \cite{Yiming:11}).  

Finally, the nodal displacement vector $\mathbf{U}^{(e)}_{pseudo-Q9}$ and the matrix $\mathbf{B}^{(e)}_{pseudo-Q9}$ are defined as follows:
\begin{equation}
\mathbf{U}^{(e)}_{pseudo-Q9}=\left[ {\begin{array}{c}
\mathbf{U}^{(e)}\\
\boldsymbol{\zeta}^{(e)}\\
	\end{array} } \right],\ \ \ 
\mathbf{B}^{(e)}_{pseudo-Q9}=\left[\mathbf{B}^{(e)}\ \ \mathbf{B}^{(e)}_\zeta\right].
\label{eq:pseudoQ9}
\end{equation} 
They are similar to the displacement vector and the $\mathbf{B}$ matrix of the Q9 element.  The original displacement of the center node of this element has been replaced by the element-wise crack openings $\zeta_n^{(e)}$ and $\zeta_t^{(e)}$.

By introducing $\Delta\Delta\boldsymbol{\zeta}^{(e)}$ as a global unknown quantity, the relationship between $\Delta\Delta \mathbf{U}^{(e)}$ and $\Delta\Delta\boldsymbol{\zeta}^{(e)}$ is established with the help of Eq.~\ref{eq:Ksub2}.  As compared to the CEM, where $\boldsymbol{\zeta}^{(e)}_{i-1}+\Delta\boldsymbol{\zeta}_{l-1}^{(e)}$ is obtained from $\mathbf{U}^{(e)}_{i-1}+\Delta\mathbf{U}_{l-1}^{(e)}$ by an additional iteration on the element level, the N-R iteration in the framework of the GCEM is more efficient, because fewer iteration steps are needed.  This will be verified in the numerical investigation.

In summary, the formulation of the cracking element, presented in \cite{Yiming:14}, is converted to matrix form, resulting in a pseudo-Q9 formulation.  With this in mind, the presented formulation can easily be changed to an enriched form of the crack opening \cite{WU2015346}.  In other words, the extra degrees of freedom for indicating crack openings are only introduced if cracks appear.

\subsection{Initiation and propagation of cracks}
The local criterion, presented in \cite{Yiming:14}, is used for determining the orientation $\mathbf{n}^{(e)}$ of the crack as
\begin{equation}
\widehat{\boldsymbol{\varepsilon}}^{(e)}\cdot\mathbf{n}^{(e)}-\widehat{{\varepsilon}}_1^{(e)}\cdot\mathbf{n}^{(e)}=\mathbf{0},
\label{eq:local}
\end{equation}
where $\widehat{\boldsymbol{\varepsilon}}^{(e)}=\sum^{8}\limits_{i=1}\left(\nabla N^{(e)}_i \otimes \mathbf{u}_i\right)^S$ and $\widehat{{\varepsilon}}_1^{(e)}$ is the first eigenvalue of $\widehat{\boldsymbol{\varepsilon}}^{(e)}$, i.e., $\mathbf{n}^{(e)}$ coincides with the unit eigenvector corresponding to $\widehat{{\varepsilon}}_1^{(e)}$\footnote{Using the Voigt notation, $\widehat{\boldsymbol{\varepsilon}}^{(e)}=\mathbf{B}^{(e),1}\ \mathbf{U}^{(e)}$ is obtained.  However, $\widehat{\boldsymbol{\varepsilon}}^{(e)}$ in Eq.~\ref{eq:local} is written in matrix form.}.  For 2D analysis, with
\begin{equation}
\widehat{\boldsymbol{\varepsilon}}^{(e)}=\left[\begin{array}{cc}
\widehat{\varepsilon}_x^{(e)}&\widehat{\gamma}_{xy}^{(e)}\ / \ 2\\
\widehat{\gamma}_{xy}^{(e)}\ / \ 2&\widehat{\varepsilon}_y^{(e)}
\end{array}\right],
\label{eq:ep2D}
\end{equation}
$\widehat{{\varepsilon}}_1^{(e)}$ is obtained as 
\begin{equation}
\widehat{{\varepsilon}}_1^{(e)}=\frac{\widehat{{\varepsilon}}_x^{(e)}+\widehat{{\varepsilon}}_y^{(e)}+\sqrt{\left(\widehat{{\varepsilon}}_x^{(e)}-\widehat{{\varepsilon}}_y^{(e)} \right)^2+\left(\widehat{\gamma}_{xy}^{(e)}\right)^2}}{2},
\label{eq:ep1}
\end{equation}
which permits determination of $\mathbf{n}^{(e)}$ based on Eq.~\ref{eq:local}.

One disadvantage of Eq.~\ref{eq:local} is that $\mathbf{n}^{(e)}$ rotates during loading.  This is even the case within one loading step if, e.g., a new crack appears.  However, since $\mathbf{n}^{(e)}$ depends on $\widehat{\boldsymbol{\varepsilon}}^{(e)}$ but not $\bar{\boldsymbol{\varepsilon}}^{(e)}$, this rotation is very small.

\begin{figure}[htbp]
	\centering
	\includegraphics[width=0.95\textwidth]{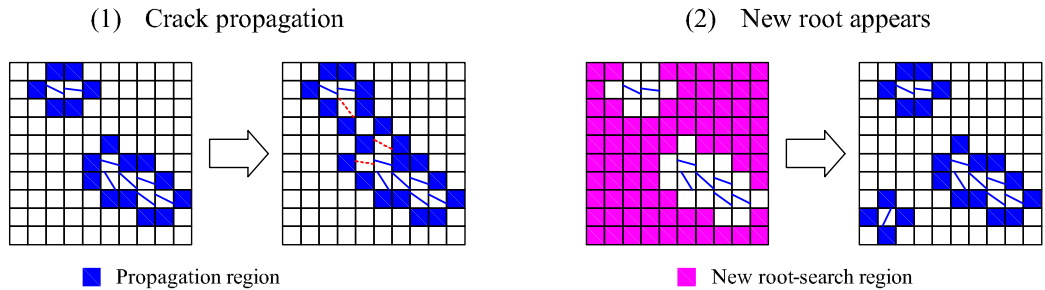}
	\caption{Separation of the uncracked elements into the propagation region and the new root-search region.  If a new cracking element appears, the propagation region will expand.}
	\label{fig:prop}
\end{figure}

Similar to the procedure described in \cite{Yiming:14}, the uncracked elements of the domain are separated into two regions: the propagation region and the new root-search region.  This is done with a simple criterion, checking whether the element shares at least one boundary with a cracking element, see Figure~\ref{fig:prop}.  The following strategy is used to find the next cracking element:

\begin{equation}
\begin{aligned}
&\mbox{find }\ \mbox{max}\left\{\phi_{RK}^{(e)}\right\} \mbox{, with} \\
&\phi_{RK}^{(e)}=\left(\mathbf{n}^{(e)}\otimes\mathbf{n}^{(e)}\right):\mathbb{C}^{(e)}:\widehat{\boldsymbol{\varepsilon}}^{(e)}-f_t^{(e)}\\
&\mbox{and }\phi_{RK}^{(e)}>0,
\label{eq:phiRK}
\end{aligned}
\end{equation}
where the propagation region is always at first searched for the existence of a propagation region\footnote{For a domain without cracks, a propagation region  obviously does not exist.  Thus, the whole domain belongs to the new root-search region.}.  If a new cracking element appears in the propagation region, this region will consequently expand, and the search will be continued.  If no new cracking element is found in the propagation region, the new root-search region will be checked.  If that a new cracking element appears in the new root-search region, the propagation region will further expand and the search in the propagation region will again be continued.  If still no new cracking element is found in the new root-search region, the next loading step will start.  With this strategy, both the initiation and the propagation of cracks can be captured.

\begin{figure}[htbp]
	\centering
	\includegraphics[width=0.75\textwidth]{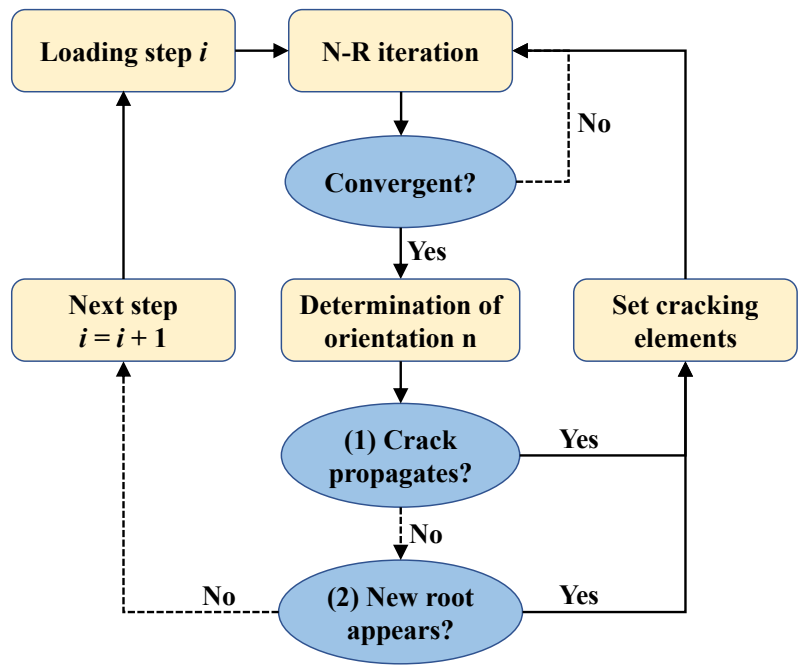}
	\caption{Calculation procedure within one N-R iteration step}
	\label{fig:procedure}
\end{figure}

The algorithm of the described procedure is illustrated in Figure~\ref{fig:procedure}.  The procedure is simpler than the one used in the framework of the CEM \cite{Yiming:14}.  The domain is firstly discretized by Q9 elements.  Once a new crack appears, the standard Q9 element is converted to a pseudo-Q9 cracking element, as described in the previous section.  

As mentioned before, the degrees of freedom of the center node of the Q9 element are ``borrowed" for indicating element-wise crack openings.  Therefore, at least theoretically these degrees of freedom can be returned.  However, the criterion is very strict insofar as only in case of $\zeta_{eq}^{(e)}=0$ and if the cracking element is not neighbor of any other cracking elements, the mentioned degrees of freedom will be returned to the center node, indicating its displacement.  In all of our numerical examples, no degrees of freedom were returned.  In other words, in all of the numerical examples the transformation from the standard Q9 to the pseudo-Q9 cracking element was never followed by its reversal.

\section{Numerical examples}
Plane-stress assumptions hold for all the numerical examples presented in this section.  
\label{sec:nurm}

\subsection{L-shaped panel}
\begin{figure}[htbp]
	\centering
	\includegraphics[width=0.85\textwidth]{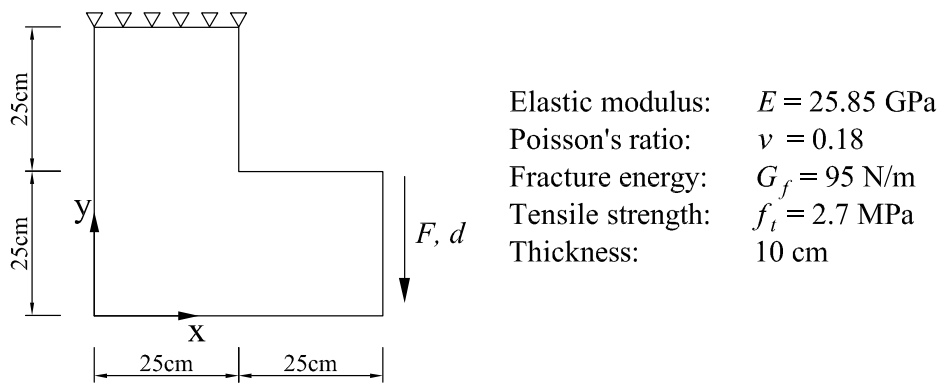}
	\caption{Set-up and material parameters of L-shaped panel test}
	\label{fig:LModel}
\end{figure}

\begin{figure}[htbp]
	\centering
	\includegraphics[width=0.99\textwidth]{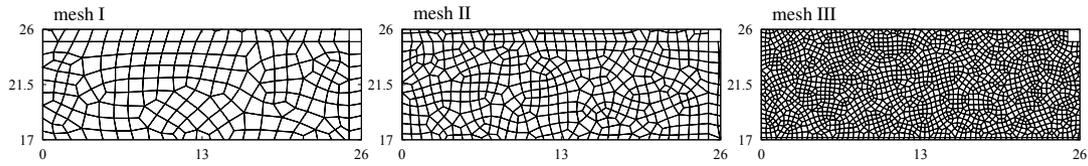}
	\caption{Meshes for the L-shaped panel test (unit: cm)}
	\label{fig:Lmeshes}
\end{figure}

The L-shaped panel test, investigated in this paper, was studied before in \cite{Yiming:11,Yiming:14}.  The set-up and the material parameters are shown in Figure~\ref{fig:LModel}.  The displacement increment is chosen as $\Delta d=10\ \mu\mbox{m}$.  The meshes considered in this work are shown in Figure~\ref{fig:Lmeshes}.

\begin{figure}[htbp]
	\centering
	\includegraphics[width=0.99\textwidth]{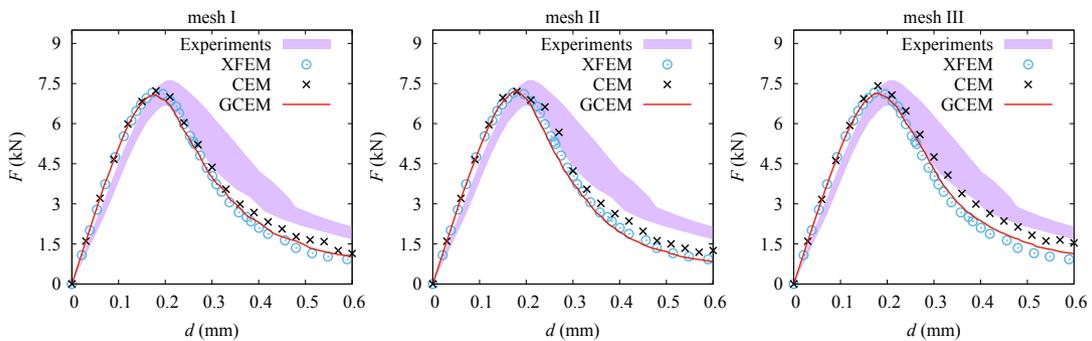}
	\caption{Force-displacement curves of the L-shaped panel for different meshes, comparing the results for the GCEM to the experimental results \cite{Winkler:02} and the results obtained by the XFEM \cite{Meschke:01} and by the CEM \cite{Yiming:14}}
	\label{fig:LForce}
\end{figure}

\begin{figure}[htbp]
	\centering
	\includegraphics[width=0.99\textwidth]{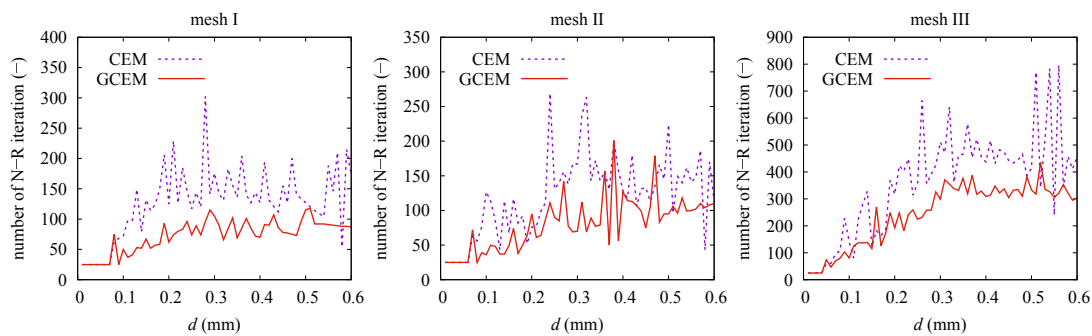}
	\caption{Comparison of the number of iterations for the GCEM with the one for the CEM}
	\label{fig:iteration}
\end{figure}

\begin{figure}[htbp]
	\centering
	\includegraphics[width=0.99\textwidth]{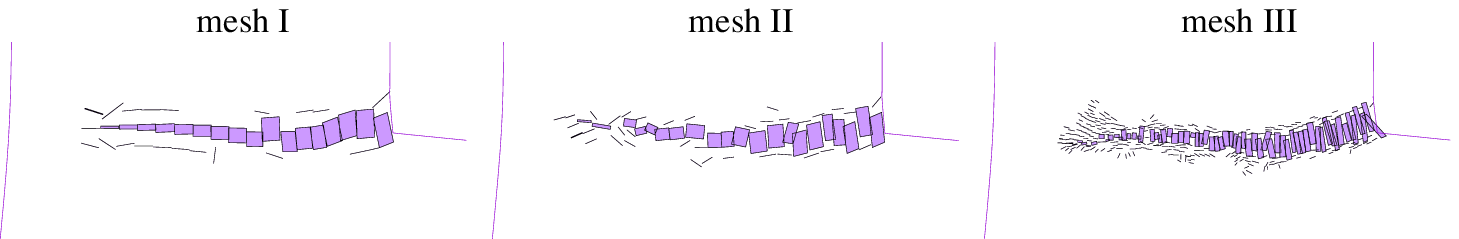}
	\caption{Deformation and crack-opening plots of the L-shaped panel, for $d=450\ \mu\mbox{m}$ (scale=1:100)}
	\label{fig:LOpening45}
\end{figure}

\begin{figure}[htbp]
	\centering
	\includegraphics[width=0.99\textwidth]{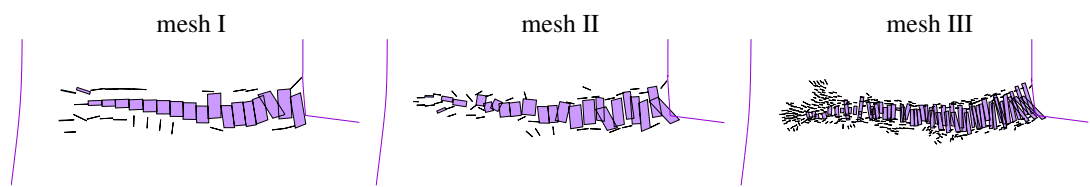}
	\caption{Deformation and crack-opening plots of the L-shaped panel, for $d=600\ \mu\mbox{m}$ (scale=1:100)}
	\label{fig:LOpening60}
\end{figure}

Force-displacement curves are shown in Figure~\ref{fig:LForce}, indicating good agreement with the results obtained by the XFEM \cite{Meschke:01} and by the CEM.  In the calculations, the total elastic energy, given in Eq.\ref{eq:Eg}, was used for checking whether the iteration converges.  Thus, if
\begin{equation}
\mbox{if } \left|\frac{\sum\limits_e E^{(e)}_l- \sum\limits_e E^{(e)}_{l-1}}{\sum\limits_e E^{(e)}_l} \right|<10^{-5},
\label{eq:converg}
\end{equation}
then the N-R iteration converged at step $l$ for the CEM and the GCEM.  The numbers of the N-R iterations are shown in Figure~\ref{fig:iteration}.  It is seen that the GCEM needs fewer iterations than the CEM.  Deformation and crack-opening plots are shown in Figures~\ref{fig:LOpening45} and~\ref{fig:LOpening60}.  The obtained results for mixed mode crack openings, for different meshes, compare well with each other.

\subsection{Brazilian disk tests with single or double slots}
Brazilian disk tests with single or double slots were studied experimentally in \cite{HAERI201420}.  Numerical simulations were conducted by the phase field method \cite{Zhou2019}, the peridynamics theory \cite{ZHOU2016235}, and by the lattice element model \cite{JIANG201741}.  Similar to the GCEM, these methods do not need crack tracking.  However, it will be shown that in case of the GCEM, good results are obtained with coarser meshes.

\begin{figure}[htbp]
	\centering
	\includegraphics[width=0.9\textwidth]{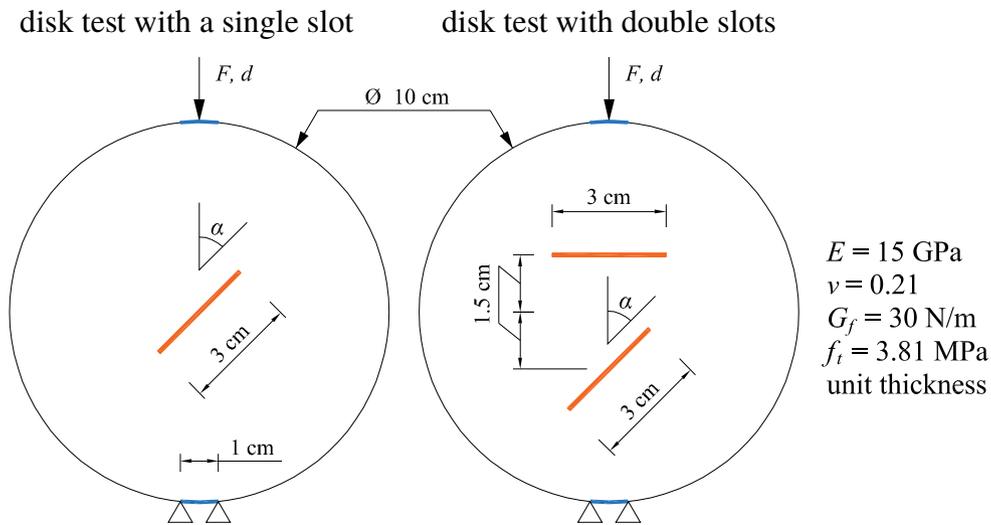}
	\caption{Model of Brazilian disk tests with single or double slots}
	\label{fig:Diskmodel}
\end{figure}

\begin{figure}[htbp]
	\centering
	\includegraphics[width=0.99\textwidth]{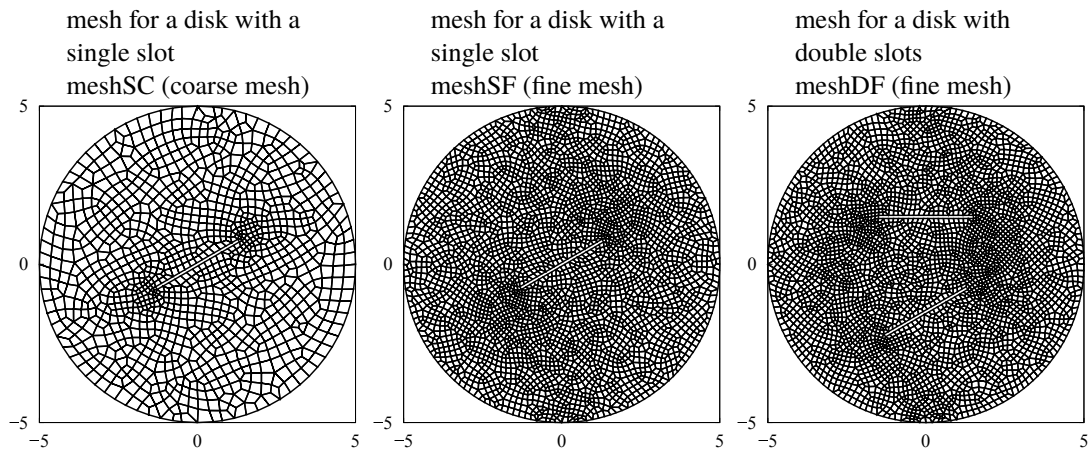}
	\caption{Meshes for Brazilian disk tests with single or double slots (unit: cm)}
	\label{fig:Diskmeshes}
\end{figure}

The models of the tests are shown in Figure~\ref{fig:Diskmodel}.  The slot width is chosen as 1~mm.  The displacement increment is chosen as $\Delta d=0.5\ \mu\mbox{m}$.    The meshes are shown in Figure~\ref{fig:Diskmeshes}.  They change slightly with different analysis of inclination, $\alpha$, of the crack.  For simulations of disk tests with a single slot, coarse as well as fine meshes are considered, whereas only fine meshes are used for simulating disk tests with double slots.  For intact disks, the relationship between the ultimate loading per unit thickness, $F_{max}$, and the tensile strength $f_t$ is given as
\begin{equation}
F_{max}=\frac{\pi \ D\ f_t}{2}=598.47~\mbox{kN},
\label{eq:Fdsplitting}
\end{equation}
which is used for determination of normalized peak loads.

\begin{figure}[htbp]
	\centering
	\includegraphics[width=0.99\textwidth]{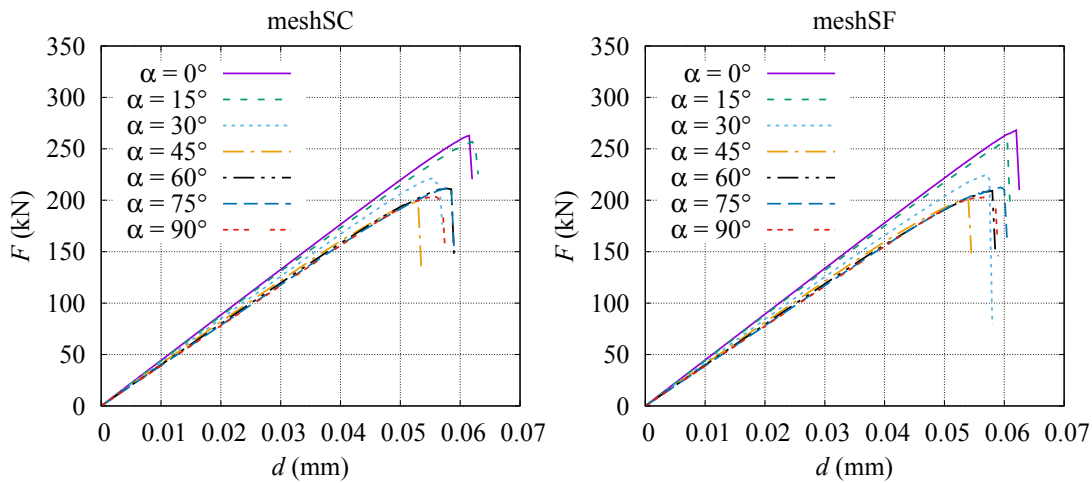}
	\caption{Force-displacement curves of disk tests with a single slot for different meshes}
	\label{fig:DiskSForce}
\end{figure}

\begin{figure}[htbp]
	\centering
	\includegraphics[width=0.99\textwidth]{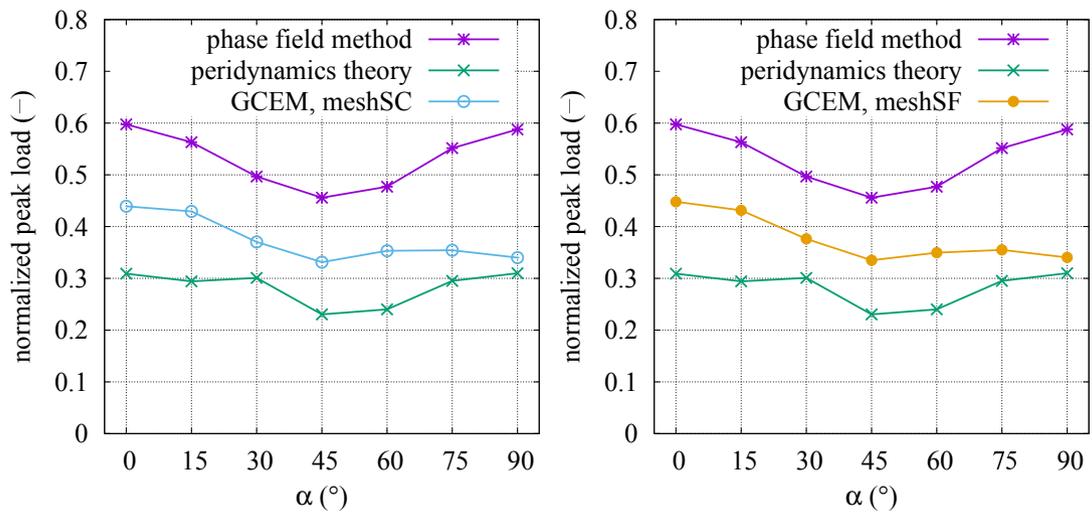}
	\caption{Comparison of normalized peak loads of disk tests with a single slot, for different meshes, with the results published in \cite{Zhou2019,ZHOU2016235}}
	\label{fig:DiskSForceNor}
\end{figure}

\begin{figure}[htbp]
	\centering
	\includegraphics[width=0.99\textwidth]{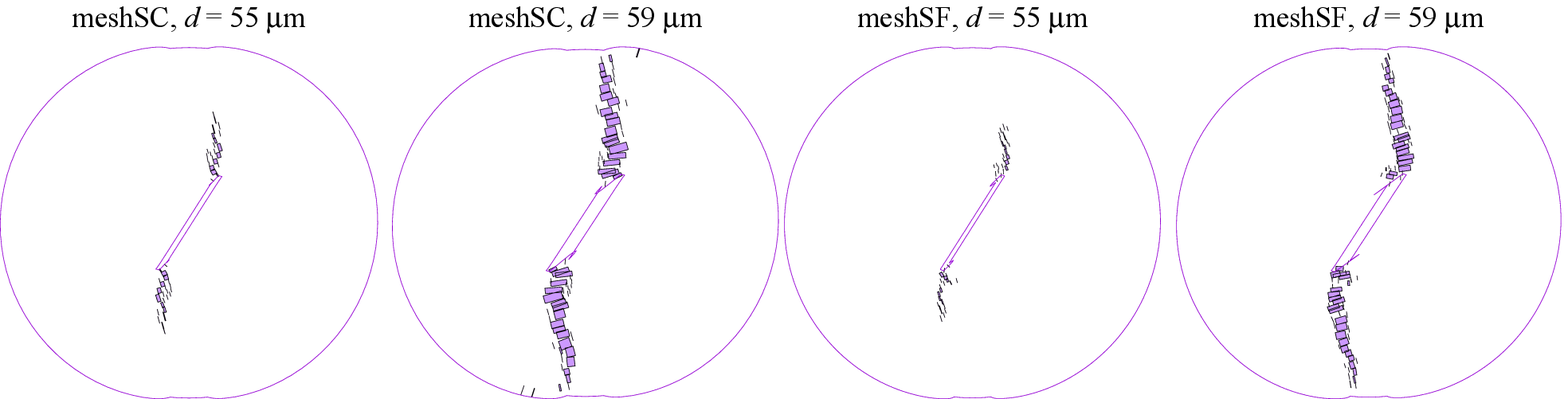}
	\caption{Deformation and crack-opening plots of disk tests with a single slot and $\alpha=30^\circ$ (scale=1:100)}
	\label{fig:DiskSOpening30deg}
\end{figure}

\begin{figure}[htbp]
	\centering
	\includegraphics[width=0.99\textwidth]{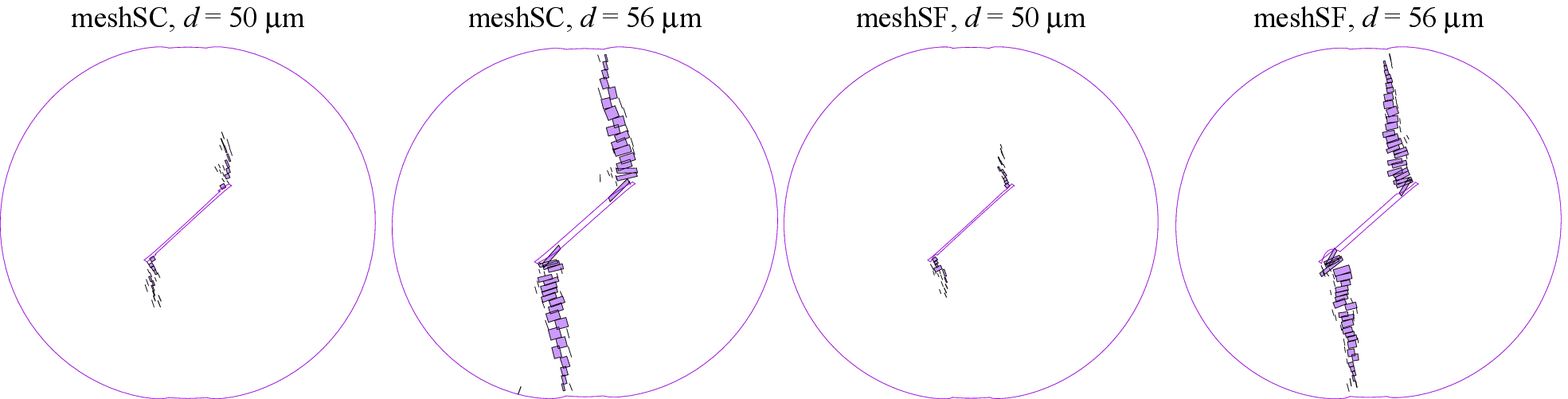}
	\caption{Deformation and crack-opening plots of disk tests with a single slot and $\alpha=45^\circ$ (scale=1:100)}
	\label{fig:DiskSOpening45deg}
\end{figure}

\begin{figure}[htbp]
	\centering
	\includegraphics[width=0.99\textwidth]{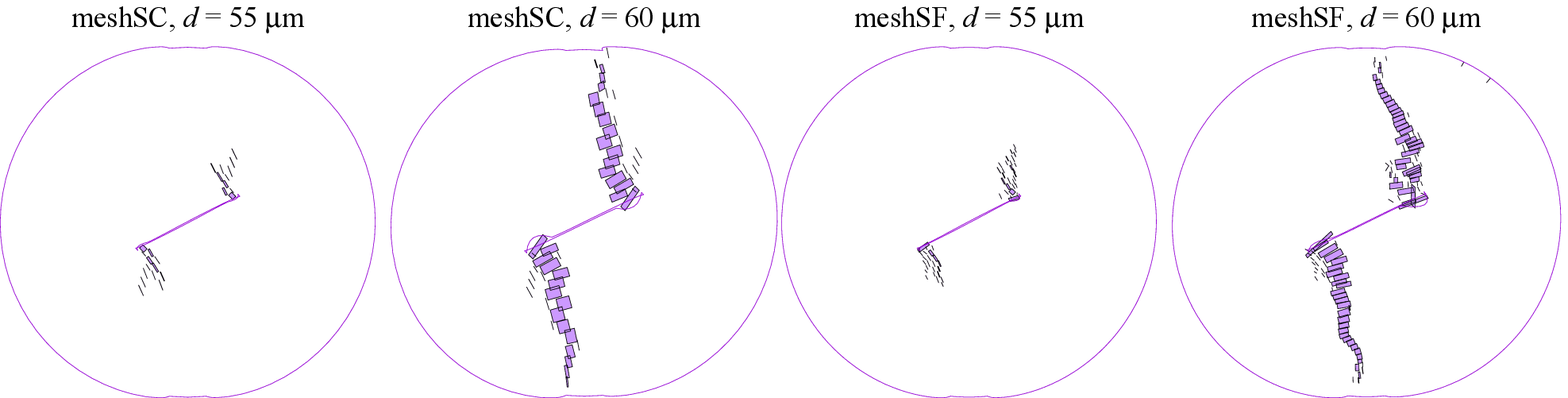}
	\caption{Deformation and crack-opening plots of disk tests with a single slot and $\alpha=60^\circ$ (scale=1:100)}
	\label{fig:DiskSOpening60deg}
\end{figure}

For disk tests with a single slot, the force-displacement curves are shown for different meshes in Figure~\ref{fig:DiskSForce}, indicating similar peak loads and load histories.  The normalized peak loads are shown in Figure~\ref{fig:DiskSForceNor} for the GCEM, the phase field method \cite{ZHOU2016235} and the peridynamics theory \cite{ZHOU2016235}.  The results provide evidence of the reliability of GCEM.  The mesh dependency is very small.  The results of the crack openings are shown in Figures~\ref{fig:DiskSOpening30deg} to~\ref{fig:DiskSOpening60deg}.  The GCEM yields element-wise disconnected crack openings.  Both the initiation and the propagation of the cracks are captured.  Moreover, not only the major cracks, but also some minor cracks are captured, see the results for $\alpha=60^\circ$ in Figure~\ref{fig:DiskSOpening60deg}. 

\begin{figure}[htbp]
	\centering
	\includegraphics[width=0.9\textwidth]{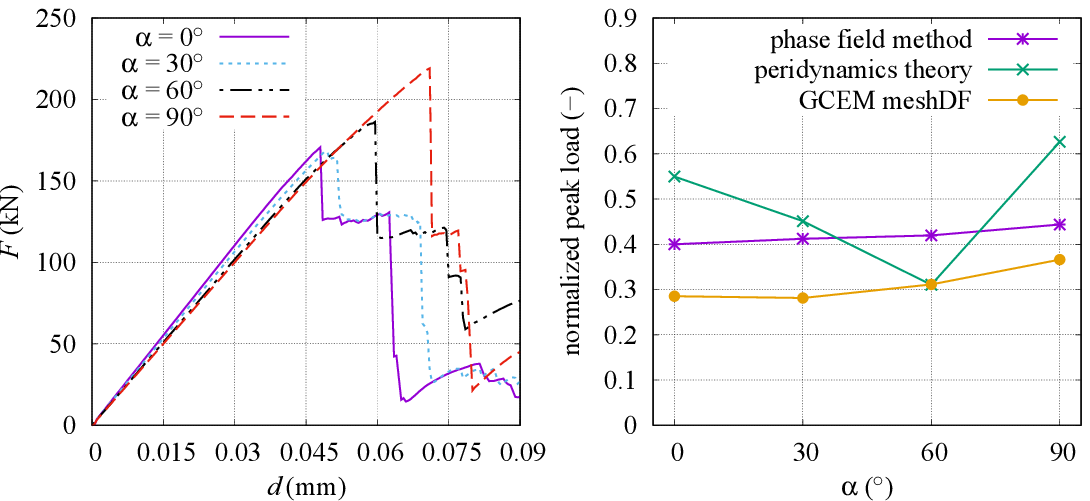}
	\caption{Comparison of results of disk tests with double slots with results reported in \cite{Zhou2019,ZHOU2016235}: (a) force-displacement curves, (b) normalized peak loads}
	\label{fig:DiskDForceRE}
\end{figure}

\begin{figure}[htbp]
	\centering
	\includegraphics[width=0.9\textwidth]{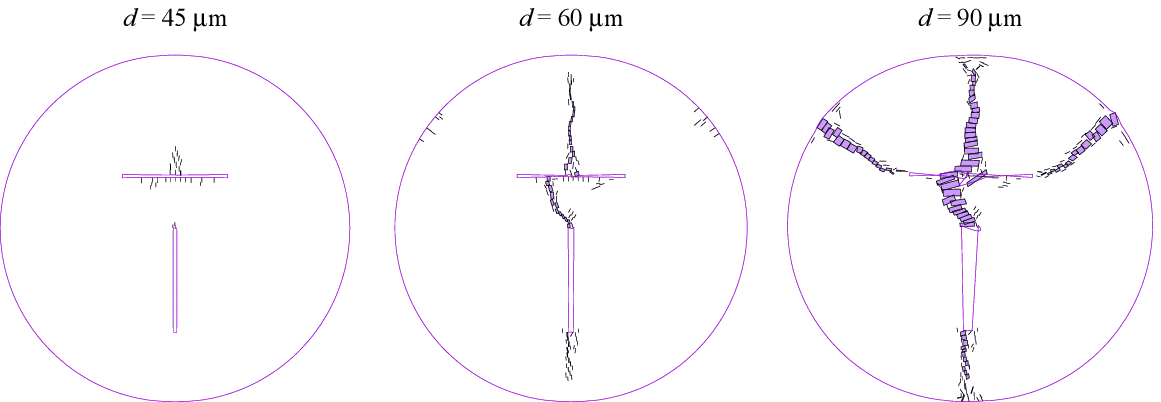}
	\caption{Deformation and crack-opening plots of disk tests with double slots and $\alpha=0^\circ$ (scale=1:15)}
	\label{fig:DiskDOpening0}
\end{figure}

\begin{figure}[htbp]
	\centering
	\includegraphics[width=0.9\textwidth]{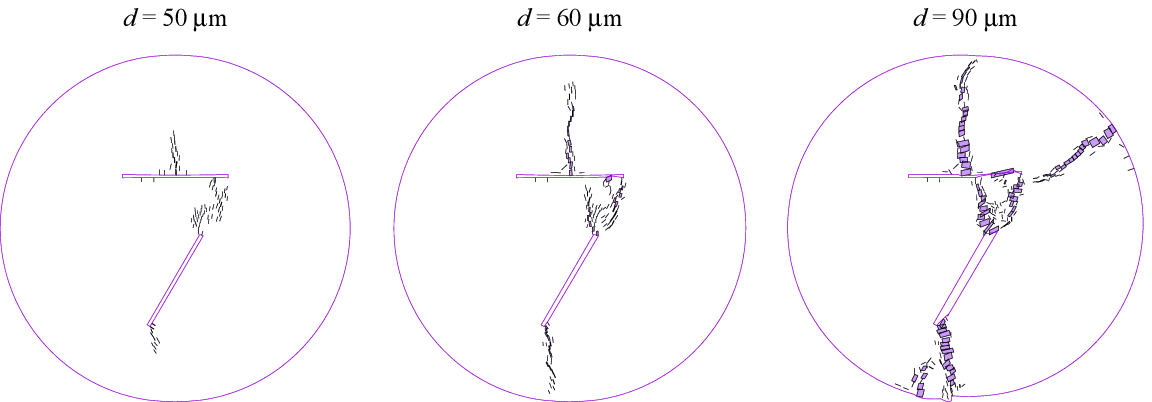}
	\caption{Deformation and crack-opening plots of disk tests with double slots and $\alpha=30^\circ$ (scale=1:15)}
	\label{fig:DiskDOpening30}
\end{figure}

\begin{figure}[htbp]
	\centering
	\includegraphics[width=0.9\textwidth]{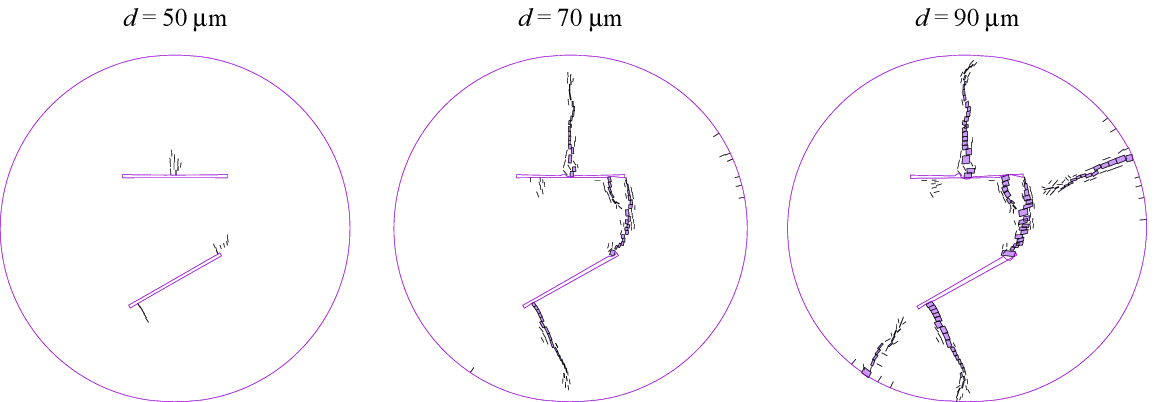}
	\caption{Deformation and crack-opening plots of disk tests with double slots and $\alpha=60^\circ$ (scale=1:15)}
	\label{fig:DiskDOpening60}
\end{figure}

\begin{figure}[htbp]
	\centering
	\includegraphics[width=0.9\textwidth]{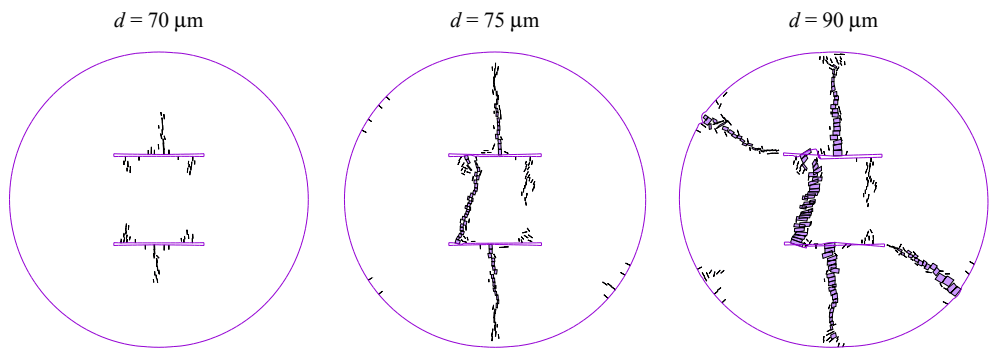}
	\caption{Deformation and crack-opening plots of disk tests with double slots and $\alpha=90^\circ$ (scale=1:15)}
	\label{fig:DiskDOpening90}
\end{figure}

For disk tests with double slots, force-displacement curves and normalized peak loads are shown in Figure~\ref{fig:DiskDForceRE}.  Generally, the latter are similar to the results published in \cite{Zhou2019}.  The results for the crack openings are shown in Figures~\ref{fig:DiskDOpening0}-\ref{fig:DiskDOpening90}, indicating that the GCEM is capable of capturing complex crack growth.

\section{Conclusions}
\label{sec:conc}
In this paper, following earlier work on the Cracking Elements Method (CEM), a novel standard Galerkin-based numerical approach, named Global Cracking Elements Method (GCEM) was presented. The Global Cracking Elements (GCE) used in the GCEM represent a special type of elements, formally similar to Q9 elements.  This feature makes it easy to implement the elements into the FEM.  Moreover, unlike the conventional XFEM and SDA, the GCEM does not need nodal/element-wise enrichment or a crack-tracking strategy.  Unlike the phase field method, a relatively coarse mesh can be used in the GCEM to obtain reliable results with negligible mesh dependency.  Unlike methods based on equivalent-type theories.  The GCEM does not introduce bonds/lattices/links and accompanied additional assumptions.  It stays in the conventional continuum-based framework.  The numerical tests have indicated the reliability, efficiency, and robustness of the GCEM.  This method is capable of capturing both the initiation and the propagation of cracks and of obtaining reliable results of mixed-mode crack openings.  

One disadvantage of the GCEM is that presently, only quadrilateral elements with nonlinear interpolation of the displacement field can be used.  Such elements are obviously more complex than elements with linear interpolation of the displacement field. The other disadvantage of the GCEM is that it avoids and abandons precise descriptions of crack tips.  The obtained crack openings are element-wise quantities.  Hence, the actual crack tips cannot be tracked directly.  Nevertheless, the GCEM is capable of capturing multiple crack growth.  This is a very difficult task even for methods that are capable of tracking the actual crack tips.

\section{Acknowledgement}
The authors gratefully acknowledge financial support by the National Natural Science Foundation of China (NSFC) (51809069) and by the Hebei Province Natural Science Fund E2019202441.



\bibliographystyle{ieeetr}
\bibliography{Reference}







\end{document}